\documentclass[11pt,a4paper]{article}
\usepackage{jheppub}
\usepackage{physics}

\usepackage{amssymb,amsfonts}
\usepackage{amssymb} 
\usepackage{amsmath, amsthm} 
\usepackage{amsgen, amsfonts, amsbsy} 
\usepackage{mathrsfs} 
\usepackage{color}

\newcommand{\be}{\begin{equation}}
\newcommand{\ee}{\end{equation}}

\definecolor{pinegreen}{rgb}{0.0, 0.47, 0.44}

\def\theequation{\arabic{section}.\arabic{equation}}

\title{\textbf{Scalar field as a perfect fluid: 
thermodynamics of minimally coupled scalars and  
Einstein frame scalar-tensor gravity}}

\author[a]{Valerio Faraoni,}
\author[b,c]{Serena Giardino,}
\author[d]{Andrea Giusti,}
\author[a]{Robert Vanderwee}

\affiliation[a]{Department of Physics \& Astronomy, Bishop's University, 
2600 College Street, Sherbrooke, Qu\'ebec, Canada J1M~1Z7}

\affiliation[b]{Max Planck Institute for Gravitational Physics (Albert 
Einstein Institute), Callinstra{\ss}e 38, 30167 Hannover, Germany}

\affiliation[c]{Institute for Theoretical Physics, Heidelberg 
University, Philosophenweg 16, 69120 Heidelberg, Germany}

\affiliation[d]{Institute for Theoretical Physics, ETH Zurich, 
Wolfgang-Pauli-Strasse 27, 8093 Zurich, Switzerland}

\emailAdd{vfaraoni@ubishops.ca}
\emailAdd{serena.giardino@aei.mpg.de}
\emailAdd{agiusti@phys.ethz.ch}
\emailAdd{rvanderwee20@ubishops.ca}

\abstract{We revisit the analogy between a minimally coupled scalar field 
in general relativity and a perfect fluid, correcting previous 
identifications of effective temperature and chemical potential.   This 
provides 
a useful complementary picture for the first-order thermodynamics of 
scalar-tensor gravity, paving the way for the Einstein frame formulation 
(which eluded previous attempts) and raises interesting questions to 
further develop the analogy.}


\arxivnumber{}

\begin{document}
\maketitle
\flushbottom

\section{Introduction}
\label{sec:1}
\setcounter{equation}{0}

Minimally coupled scalar fields are ubiquitous in theoretical physics and 
especially in cosmology. One of the simplest examples is quintessence, a 
proposal of dynamical dark energy represented by a scalar field, that aims 
to avoid the fine-tuning issues of a cosmological constant and could 
explain the current accelerated expansion of the universe in the context 
of general relativity (GR) \cite{AmendolaTsujikawa,Peebles:2002gy}. Also 
more elaborated scalar field models with non-canonical kinetic term are 
considered for dark energy, {\em e.g.}, in the so-called k-essence 
\cite{Armendariz-Picon:1999hyi, Armendariz-Picon:2000ulo, Bilic:2008zk}. 
Other exotic models of scalar-field based quintessence include, {\em 
e.g.}, the string-inspired tachyon condensate \cite{Padmanabhan:2002sh, 
Sen:2002an, Sen:2002nu, Sen:2002in, Gibbons:2003gb}.

The equivalence between a scalar field minimally coupled to the curvature 
endowed with timelike gradient and a perfect fluid is now well-established 
and has been the subject of a substantial literature 
(\cite{Unnikrishnan:2010ag,Faraoni:2012hn, Semiz:2012zz, Diez-Tejedor:2005oom, 
Diez-Tejedor:2013nwa,Piattella:2013wpa,Gergely:2020ncx} and references 
therein). The 
generalization of this equivalence to imperfect fluids has proven to be 
even more fruitful, especially for applications to dark energy (and more 
general dark sector) models. One of the most intriguing developments 
involving imperfect fluids is the class of scalar-tensor theories that 
contain an essential mixing of scalar and tensor kinetic terms known as 
Kinetic Gravity Braiding 
\cite{Deffayet:2010qz,LimSawickiVikman10,Pujolas:2011he,Mirzagholi:2014ifa}, leading to a rich 
dark energy phenomenology. The imperfect fluid description in these works 
is based on the notion of chemical potential \cite{Pujolas:2011he}, which 
we are interested in here.

Naturally, since the main interest in the fluid description of minimally 
coupled scalar fields arises in cosmology, the theory of cosmological 
perturbations generated by scalar field matter is addressed in numerous works ({\em e.g.}, \cite{Khlopov:1985jw, 
Garriga:1999vw, 
Fabris:2000ee, Arroja:2010wy, Sawicki:2012re, Christopherson:2012kw} 
and references therein). 
Moreover, the effective field theory of relativistic media, 
including fluids, solids and exotic hypothetical media, has been developed 
in \cite{Dubovsky:2005xd, Dubovsky:2011sj, Nicolis:2015sra, 
Ballesteros:2016gwc, Nicolis:2013lma, Ballesteros:2014sxa, 
Ballesteros:2016kdx}.

The fluid description of minimally coupled scalar fields crucially relies 
on the scalar field gradient being timelike, in order to be able 
to meaningfully 
define a fluid four-velocity. The analysis of scalars with 
non-timelike 
gradients has also been attempted \cite{Semiz:2012zz, Gergely:2020ncx}, 
but is still in its infancy.

In this work, we revisit the analogy between perfect fluid and minimally 
coupled scalar field which, albeit well-known, still leaves room for 
interesting developments. More specifically, a thermodynamical description 
for this fluid was recently presented \cite{Piattella:2013wpa}, 
introducing the notions of temperature and chemical potential for the 
fluid. However, these results pose problems that we aim to 
solve. Addressing these issues also makes it possible 
to understand the analogy 
in a broader picture, by connecting it with the more 
general thermodynamical 
description of imperfect fluids in the context of scalar-tensor gravity \cite{Faraoni:2021lfc, 
Faraoni:2021jri,Giardino:2022sdv,Faraoni:2022jyd}. This link additionally 
allows one to shed light on the Einstein frame formulation of first-order 
thermodynamics, which has so far remained elusive.

We first review some generalities to set the stage for our analysis.
The action of gravity with a minimally coupled scalar field is 
\be
\label{min}
S=\int d^4 x \, \sqrt{-g} \left[ \frac{R}{2\kappa} +
{\cal L}\left( \phi, X \right) 
\right] +  S^\mathrm{(m)} \,,
\ee
where\footnote{We follow the notation of Ref.~\cite{Waldbook}. The
metric signature is $({-}{+}{+}{+})$ and units are used in which Newton's 
constant $G$ and the speed of light $c$ are unity.}   $g$ is 
the determinant of the metric tensor $g_{ab}$, $R$ is the 
Ricci scalar, $\kappa=8\pi G$, $G$ is Newton's constant, 
\be
X \equiv -\frac{1}{2} \, \nabla^c \phi \nabla_c \phi \,,
\ee
${\cal L}\left( \phi, X \right)$ is the scalar field Lagrangian density, 
and $ S^\mathrm{(m)}$ describes matter other than the scalar 
field.

In the rest of this paper we assume that the scalar field has timelike 
gradient, $X>0$. As is well-known ({\em e.g.}, \cite{Madsen:1988ph, 
Pimentel89,Faraoni:2012hn, Semiz:2012zz, Faraoni:2018qdr,Quiros:2019gai, 
Giusti:2021sku}), one can then  establish an analogy between the scalar 
and a perfect fluid by taking the normalized gradient of $\phi$ as the  
fluid's four-velocity,
\be
u^a \equiv \frac{\nabla^a \phi}{\sqrt{2X}} \label{4-velocity}
\ee
(the sign of this vector must be adjusted to make it future-oriented). 

Several works have derived the fluid-mechanical quantities corresponding 
to the minimally coupled scalar field \cite{Unnikrishnan:2010ag, 
Deffayet:2010qz, LimSawickiVikman10, Pujolas:2011he, Faraoni:2012hn, 
Semiz:2012zz, Mirzagholi:2014ifa, Diez-Tejedor:2013nwa, Piattella:2013wpa, 
Gergely:2020ncx} and its thermodynamics has been studied recently in 
\cite{Piattella:2013wpa}. The effective $\phi$-fluid is a perfect fluid 
characterized by its four-velocity $u^a$, effective energy density $\rho$, 
pressure $P$, number density $n$, entropy density $s$, temperature ${\cal 
T}$, and chemical potential $\mu$.  The current state of knowledge of the 
 scalar field-fluid correspondence is summarized by  
Eq.~(\ref{4-velocity}) and by the 
following dictionary appearing in Table~1 of 
\cite{Piattella:2013wpa}:
\begin{eqnarray}
\rho & = & 2X {\cal L}_X- {\cal L} \,, \label{5}\\
&&\nonumber\\
P & = & {\cal L} \,,\label{6} \\
&&\nonumber\\
n & = & \sqrt{2X}\,  {\cal L}_X \label{7} \,,\\
&&\nonumber\\
\frac{s}{n} & = & \phi \,,\label{8} \\
&&\nonumber\\
{\cal T} & = & \frac{ - {\cal L}_{\phi} }{ \sqrt{2X} \, {\cal L}_X}  
\,,\label{9} 
\\
&&\nonumber\\
\mu & = & \frac{ 2X {\cal L}_X+\phi {\cal L}_{\phi} }{\sqrt{2X} \, {\cal 
L}_X } = \sqrt{2X}  -\phi {\cal T}  \,, \label{10} 
\end{eqnarray}
where ${\cal L}_{\phi} \equiv \partial {\cal L}/\partial \phi$ and ${\cal L}_X \equiv \partial {\cal L}/\partial X$. 
Assuming ${\cal L}_X>0$, {\em i.e.}, that the field $\phi$ is not a 
phantom, the energy density $\rho$ and the particle number density $n$ are 
non-negative. Using Eqs.~(\ref{4-velocity}), (\ref{5}), and (\ref{6}), 
the stress-energy tensor of the scalar field
\be
T_{ab}^{(\phi)} = {\cal L}_X \nabla_a \phi \nabla_b \phi +{\cal L} g_{ab}  
\,,
\label{phiTab}
\ee
which is conserved ($\nabla^b T_{ab}^{(\phi)}=0$), is rewritten in the 
perfect 
fluid form
\be
T_{ab}^{(\phi)} = 
\left( \rho +P\right)u_a u_b +P g_{ab} =\rho u_a u_b +P h_{ab}
\,, \label{11}
\ee
where $h_{ab} \equiv g_{ab} +u_a u_b$ is the Riemannian metric on the 
3-space orthogonal to $u^a$, satisfying
\be
h_{ab} u^a =h_{ab} u^b=0
\ee
(${h^a}_b$ is the projection operator onto this 3-space). 

The equation of motion for $\phi$ 
\be
\nabla_a \Big( {\cal L}_X  \nabla^a \phi \Big) = -{\cal L}_{\phi} 
\ee
is written as 
\be
\nabla_a \Big( nu^a \Big)\equiv \nabla_a N^a =-{\cal L}_{\phi} \,,
\ee
which reduces to the familiar Klein-Gordon equation $\Box \phi=V_\phi $ 
if ${\cal L}\left( 
\phi, X \right) = X-V(\phi)$, where $V(\phi)$ is the scalar field 
potential.

If ${\cal L}={\cal L}(X)$,  the scalar field theory is 
invariant under the shift 
symmetry $\phi \to \phi +C$ (where $C$ is a constant) and there is a conserved 
Noether current 
\be 
N^a = {\cal L}_X \nabla^a \phi = n u^a 
\ee 
satisfying 
\be
\label{ncons}
\nabla_a N^a=0 \,. 
\ee
$N^a$ is the analogue of the particle number current density. The particle 
number density in the comoving frame is the corresponding Noether charge 
\be 
n = -N^0 = -u^c N_c = -\frac{\nabla^a \phi}{\sqrt{2X}} \, {\cal L}_X \nabla_a \phi = 
\sqrt{2X} \, {\cal L}_X \,, 
\ee 
consistently with Eq.~(\ref{7}). If ${\cal L}_{\phi} \neq 0$ 
(for example, if there is a potential $V(\phi)$), the analogue 
\be 
N^a =n u^a 
\ee 
of the particle current density is not conserved,\footnote{In a 
dissipative fluid, the directions of the particle flow and of the energy 
flow are different. As a consequence, $N^a$ coincides with $n u^a$ in the 
comoving (or Eckart) frame \cite{Eckart40} which is adapted to follow the 
total flux of particles, while $ N^a=nu^a + v_{(L)}^a $ in the Landau (or 
energy frame) \cite{Landau}, where $v_{(L)}^a$ is the diffusive current 
density of particles caused by gradients of the chemical potential $\mu$.  
The Eckart and the Landau frames coincide for a perfect fluid, which is 
 the case we are interested in.} $\nabla_a N^a= 
-{\cal L}_{\phi} \neq 0 $. Being derived from a scalar field, the 
$\phi$-fluid is, of course, irrotational (the kinematic quantities 
associated with the fluid four-velocity are computed in 
\cite{Faraoni:2018qdr}).

Most of this analogy has been  derived and validated in several 
situations and 
appears to be a special case of the more general equivalence between 
a scalar field coupled nonminimally to the Ricci scalar $R$ and an 
effective {\em imperfect} fluid exhibiting heat conduction, bulk and shear 
viscosity, and anisotropic stresses \cite{Madsen:1988ph,
Pimentel89,Faraoni:2018qdr,Quiros:2019gai, Giusti:2021sku}. 
The stress-energy tensor of 
the effective dissipative fluid in this more general case has the form
\be
T_{ab}^\mathrm{(dissipative)} = \rho u_a u_b +P h_{ab} +q_a u_b + q_b u_a 
+\pi_{ab} \,,\label{12}
\ee
where
\be
P= P_\mathrm{non-viscous} +P_\mathrm{viscous}
\ee
is the sum of  a non-viscous and of a viscous pressure (the latter is associated with 
bulk viscosity), $q^a$ 
is a 
purely spatial ({\em i.e.}, $q_c u^c=0$) heat flux density, and $\pi_{ab}$ 
(with ${\pi^a}_a=0$, $ \pi_{ab}u^a =\pi_{ab} u^b=0$) 
denotes the anisotropic stresses. If the scalar field $\phi$ is minimally 
coupled to $R$, all the imperfect fluid terms vanish and Eq.~(\ref{12}) 
reduces to the perfect fluid form~(\ref{11}). This analogy has been 
 often studied in the context of 
FLRW cosmology and of the special theory of conformally/nonminimally 
coupled scalar fields, but has seldom been considered for general 
``first-generation'' scalar-tensor gravity 
\cite{Pimentel89,Faraoni:2018qdr}. More recently, it has been extended to 
Horndeski gravity  \cite{Quiros:2019gai, Giusti:2021sku}.

The introduction of temperature ${\cal T}$ and chemical potential $\mu$ in 
the correspondence between minimally coupled scalar field and perfect 
fluid is quite recent (appearing only in \cite{Piattella:2013wpa} to the 
best of our knowledge) and has not been tested as well as the rest of the 
analogy. Indeed, the derivation of ${\cal T}$ and, as a consequence, of 
$\mu$ in \cite{Piattella:2013wpa} exhibits an inconsistency (that does not 
affect the other fluid quantities), that we correct here. Before analysing 
the details (Sec.~\ref{sec:2}), it is already apparent that $T$ and $\mu$ 
given by Eqs.~(\ref{9}) and (\ref{10}) suffer from three 
problems.

\begin{enumerate}

\item The first issue (already noted in 
\cite{Piattella:2013wpa}) is that both ${\cal T}$ and $\mu$ can be 
negative. This fact is surprising because, contrary to the nonminimally 
coupled scalars of scalar-tensor gravity, the effective $\phi$-fluid is 
otherwise well-behaved and satisfies the weak 
and null energy conditions, hence one expects 
${\cal T}$ and $\mu$ to be non-negative like $\rho$ and $n$.

\item The most serious problem is that, according to Eq.~(\ref{9}), there 
is a temperature gradient. Moreover, in general the effective $\phi$-fluid 
is non-geodesic, with non-zero acceleration
\begin{eqnarray}
\dot{u}_a & \equiv & u^c \nabla_c u_a  
= - \frac{1}{{2 X}} \left(\nabla _a X + \frac{\nabla_c X  
\nabla^c \phi}{2 X} \, \nabla _a \phi \right)  \,. 
\label{acceleration}
\end{eqnarray}
Then, there must necessarily be a heat current with density 
\cite{Eckart40} 
\be 
q_a = - {\cal K} \left( h_{ab}\nabla^b {\cal T}  + {\cal T} \dot{u}_a \right) 
\,, \label{eq:q} 
\ee 
where $ {\cal K}$ is the (analogue of) the thermal conductivity. This 
generalized Fourier law is one of the three constitutive equations of 
Eckart's first-order thermodynamics \cite{Eckart40} and a minimal 
assumption. The first term in the right-hand side of Eq.~(\ref{eq:q}) is 
nothing but the usual Fourier law, while the second one is a purely 
relativistic ``inertial'' contribution discovered by Eckart 
\cite{Eckart40}. The heat conduction described by $q_a$ makes a fluid 
dissipative and endows its stress-energy tensor with the dissipative terms 
appearing in Eq.~(\ref{12}). Then, how can the fluid equivalent of a 
minimally coupled $\phi$ be a perfect fluid described by~(\ref{11})? A 
heat current would necessarily show up in the comoving (or Eckart) frame 
based on the four-velocity~(\ref{4-velocity}). The {\em only} way for 
$q_a$ to 
vanish identically is if $ {\cal T} = 0 $.

\item Although here we limit ourselves to scalar fields coupled minimally 
to $R$, from the perspective of the broader nonminimally coupled scalar 
field thermodynamics (which is still under development but has certain 
firm points) the fact that a minimally coupled scalar field fluid is 
endowed with a non-zero temperature appears very odd. In that context 
\cite{Faraoni:2021lfc, Faraoni:2021jri, 
Giusti:2021sku,Giardino:2022sdv,Faraoni:2022jyd}, the nonminimal coupling with $R$ is responsible for a nonvanishing fluid 
temperature, therefore the fluid equivalent to a minimally coupled $\phi$ 
and with Lagrangian depending only on $\phi$ and $X$ should always have 
zero temperature.

\end{enumerate}

In the rest of this article we address these problems. We begin by 
correcting the temperature~(\ref{9}), establishing the fact that the fluid 
equivalent to a minimally coupled $\phi$ has always zero temperature, 
resolving the first conundrum of scalar-tensor thermodynamics. As a 
consequence, only the first term in the chemical potential $ \mu = \sqrt{2X}-\phi {\cal T} $ remains, which makes this 
quantity positive-definite. The second and third issue  listed above are 
also solved because the heat flux density $q_a$ then vanishes identically 
and the fluid becomes non-dissipative.

After describing the thermodynamics of 
the fluid equivalent to a minimally coupled $\phi$, we are  also able to 
comment on the 
thermodynamics of phantom scalar fields with ${\cal L}_X<0$, which have 
been the subject of an extensive literature, in conjunction with studies 
considering the possibility of a very negative equation of state ($w 
\equiv P/\rho <-1$) for the dark energy driving the present 
acceleration of the cosmic expansion, {\em e.g.} \cite{Nesseris:2006er,DiValentino:2016hlg,DiValentino:2020naf}.
Although the claims of 
a phantom equation of state 
are disputed, the possibility of $w<-1$ is not excluded by present 
cosmological observations \cite{Planck:2018vyg}.  
Phantoms are unstable from the classical, and even more 
from the quantum, point of view but they are 
still accepted in the cosmological literature as the expression of a 
truncated action that will be cured if all terms are included. The 
literature on phantom field thermodynamics, now mostly a decade 
old, has not 
been conclusive and we contribute to a clearer picture.

Finally, we can extend the fluid-$\phi$ correspondence to include scalar 
fields coupled nonminimally {\em to matter} (but not to $R$). The effect 
of these couplings is analogous to that of a scalar field potential which 
constitutes a source of fluid ``particles'' making the ``particle number 
density'' $n$ a non-conserved quantity, but has no drastic effect on the 
rest of the analogy. This extension allows one to discuss the Einstein 
frame version of scalar-tensor gravity in which the gravitational 
Brans-Dicke-like field $\tilde{\phi}$ couples explicitly to matter but not 
to $R$ (contrary to the Jordan frame formulation of the same theory in 
which the scalar $\phi\neq \tilde{\phi}$ couples to $R$ but not to 
matter).  This development makes it possible to fill a gap in the 
first-order thermodynamics of scalar-tensor gravity which, being based on 
the notion of temperature, was thus far unable to deal with the Einstein 
frame description.

\section{Temperature and chemical potential in the scalar field-fluid 
analogy}
\label{sec:2}
\setcounter{equation}{0}

Consider the first law of thermodynamics \cite[Box 22.1, p.~561]{MTW}
\be
d \left(\frac{\rho}{n} \right)+ P \, d \left( \frac{1}{n} \right) = {\cal T} \, d \left(\frac{s}{n} \right) \, ,
\ee
where ${\cal T}$ denotes the temperature, $s$ the entropy per unit volume ({\em i.e.}, $s = S/V$),
$\rho$ is the internal energy density (per unit volume), and $P$ is the 
pressure. The symbol $s$ in \cite{MTW} 
corresponds to $s/n$ in our discussion.

Now, taking $s$ and $n$ as independent variables from \cite[Box 22.1, p.~561]{MTW} one has that
\be
{\cal T} (s, n) = \frac{1}{n} \frac{\partial \rho}{\partial (s/n)} 
\Big|_n = \frac{\partial \rho}{\partial s} \Big|_n \quad ,
\ee
thus, since Eq.~\eqref{phiTab} maps into a perfect fluid, the absence of 
any dissipative effects, hence vanishing heat fluxes, requires ${\cal T} 
= 0$, in accordance with the principles of the thermodynamics of 
scalar-tensor gravity 
\cite{Faraoni:2021lfc,Faraoni:2021jri,Giusti:2021sku,Giardino:2022sdv,Faraoni:2022jyd}. 
Therefore, assuming $\phi = \phi (s,n)$ and $X 
= X (s,n)$, one has that
\be
0 = \frac{\partial \rho}{\partial s} \Big|_n = - \mathcal{L}_{\phi} 
\,\frac{\partial \phi}{\partial s} \Big|_n \quad ,
\label{condonphi}
\ee
where we have taken advantage of Eqs.~\eqref{5} and \eqref{7}. The 
condition in Eq.~\eqref{condonphi} is then satisfied if  
$\mathcal{L}_{\phi} = 0$ or $ \frac{\partial \phi}{ \partial s} \Big|_n 
= 0$. 
Since, in general, $\mathcal{L}$ will contain a potential  term, 
consistency with the thermodynamics of scalar-tensor gravity 
\cite{Faraoni:2021lfc, Faraoni:2021jri,Giusti:2021sku,Giardino:2022sdv,Faraoni:2022jyd} 
translates into the condition 
\be
\frac{\partial \phi}{\partial s} \Big|_n = 0 \, ,
\ee
so that the temperature of gravity vanishes for a scalar field 
non-minimally coupled to Einstein gravity. 

On a similar note, it is easy to see that combining \cite[Box 22.1, 
p.~561]{MTW}
\be
P \left( n, s \right) = n\, \frac{\partial \rho}{\partial n} 
\Big|_{s/n} -\rho \, ,
\ee
with Eqs.~\eqref{5} and \eqref{6} recovers the perfect fluid 
identification $P = \mathcal{L}$ if and only if
\be
\frac{\partial \phi}{\partial n}\Big|_{s/n} = 0 \, ,
\ee
when $\mathcal{L}_\phi \neq 0$.

It is then easy to identify the chemical potential of the system, which 
reads \cite[Box 22.1, p.~561]{MTW}
\be
\mu = \frac{P + \rho}{n} = \sqrt{2X} \, ,
\ee
where we have again taken advantage of Eqs. \eqref{5}--\eqref{7}.

The condition in Eq.~\eqref{8} is incompatible with both the thermodynamic 
analogy presented here and the requirement of conservation of the entropy 
per particle along perfect fluid lines (see Appendix \ref{AppendixA}). 
This condition is, however, marginal in our discussion since it is not 
used.

\subsection{Approach to the diffusive equilibrium}

One can {\em a posteriori} derive an equation describing the approach to 
diffusive equilibrium along the fluid lines. For relativistic fluids, the 
chemical potential $\mu$ and the (purely spatial) diffusive flux density 
of particles $q_a^{(p)}$ will obey a generalization of Fick's law 
analogous to 
Eckart's generalization~(\ref{eq:q}) of Fourier's law (cf. 
Ref.~\cite{Kremer:2013fxa})
\be
q_a^{(p)} = -{\cal D} \left( h_{ab}\nabla^b \mu + \mu \, \dot{u}_a \right) 
\,,\label{generalizedFick}
\ee
where ${\cal D}$ is a diffusion coefficient analogous to the thermal 
conductivity ${\cal K}$.  Diffusive equilibrium is reached when the 
chemical potential $\mu$ vanishes identically (in the presence of 
acceleration $\dot{u}^a$, a constant $\mu$ would still generate particle 
diffusion due to the second term in the right-hand side of 
Eq.~(\ref{generalizedFick}). Equation~(\ref{generalizedFick}) is 
reminiscent of a relativistic version of the drift-diffusion equation 
\cite{Chandrasekhar:1943ws}. This is used, for example, in the context of 
semiconductors \cite{hu2015simulation}, where it describes particle 
currents (for electrons and holes) in terms of the particle density 
gradients and a term containing the electric field vector.\footnote{
The similar form of the drift current and of Eckart's heat current density 
is due to the presence of a force, hence of an acceleration, in the 
constitutive law leading to the expression of the drift current. This 
similarity could potentially be of interest for analogue gravity.}

Let us compute the derivative $d\mu/ d\tau $,  
where $\tau$ is the proper time along the flow lines of the effective
$\phi$-fluid:
\begin{eqnarray}
\frac{d\mu}{d\tau} \equiv  u^c \nabla_c \mu = 
\frac{\nabla^c \phi}{\sqrt{2X}} \, \nabla_c \left( \sqrt{2X} \right) =
\frac{\nabla^c \phi \, \nabla_c X}{2X}
 \,. \label{temp1}
\end{eqnarray}
Now use the expression of the expansion scalar of the $\phi$-fluid 
\cite{Faraoni:2018qdr,Giusti:2021sku}
\be
\Theta \equiv \nabla _a u^a = 
\frac{1}{\sqrt{2 X}} \left( \Box \phi - \frac{\nabla_c X  
\nabla^c \phi}{2 X} \right) 
\ee
to eliminate the term containing second derivatives of $\phi$ in 
Eq.~(\ref{temp1}), obtaining
\be
\frac{d\mu}{d\tau} = -\mu \, \Theta +\Box \phi \,.\label{approach}
\ee
This equation is not so simple because of the d'Alembertian of $\phi$ in 
the 
right-hand side. However, to gain some insight, we can 
consider the situation in which ${\cal L}$ does not depend on $\phi$, for 
example a  free scalar field with ${\cal L}=X$, in which case $\Box 
\phi=0$ and  Eq.~(\ref{approach}) reduces to\footnote{If ${\cal L}_X=1$, 
then $n=\mu$ satisfies the 
same equation, which is reported in \cite{Ballesteros:2016kdx}.} 
\be
\dot{\mu} = -\mu \, \Theta \,.
\ee
 One can introduce a representative 
length $\ell$ by \cite{Ellis:1971pg}
\be
\frac{ \dot{\ell}}{\ell} \equiv \frac{ \Theta}{3} 
\ee
and then 
\be
\frac{ \dot{\mu}}{\mu} = -\frac{3\dot{\ell}}{\ell} 
\ee
so that $\mu = \mbox{const.}/\ell^3$. The simplified evolution equation of 
$\mu$ then simply says that when the flow expands and dilutes, $\mu$ 
decreases 
and the state of equilibrium $\mu=0$ is approached,  while when the flow 
gets concentrated, the chemical potential increases and there is departure 
from the equilibrium state. In particular, the $\phi$-flow is  
diluted in an expanding universe, which will approach the diffusive 
equilibrium state $\mu=0$ as $\ell \to +\infty$. Near spacetime 
singularities, instead, this flow is focused, the flow lines become  
closer and closer, and there are extreme departures from the equilibrium 
state $\mu=0$.
In principle, this understanding of the approach to equilibrium in 
the thermodynamical picture based on $\mu$ is equivalent 
to that obtained in the 
context of scalar-tensor thermodynamics based on ${\cal T}$ 
\cite{Faraoni:2021lfc, Faraoni:2021jri,Giardino:2022sdv,Faraoni:2022jyd}. 
However, in the comoving frame one does not see particle 
diffusion, as explained in the next subsection. 
When the second term $\Box \phi$ is included in the right-hand side of 
Eq.~(\ref{approach}), the situation becomes more complex since this 
term could in principle be positive or negative, hence it can favour the 
approach to equilibrium or oppose it depending on its sign.

\subsection{No diffusive particle current in the comoving frame}

The effective stress-energy tensor~(\ref{phiTab})  of the effective 
$\phi$-fluid has the perfect fluid form (\ref{11}), yet the chemical 
potential $\mu= \sqrt{2X}$ 
depends on the spacetime position so its variation must give rise to a 
diffusive $\phi$- (or ``particle'') current (the acceleration also 
contributes to this diffusive flow according to  
Eq.~(\ref{generalizedFick})). 
Then it is natural to ask why we do not see a vector $q^a_\mathrm{(p)}$ 
describing this diffusion in the effective fluid stress-energy 
tensor. The answer is well-known to researchers working with relativistic 
dissipative fluids, in which the direction of the energy flow is distinct 
from that of the particle flow. In  dissipative fluids, the Eckart (or 
comoving) frame is based on the particle 
four-velocity (that is, the four-velocity of the Eckart or comoving 
observers 
is the $u^a$  of the fluid given by Eq.~(\ref{4-velocity})). Since this 
frame is adapted to follow the total flow of particles, the diffusive 
particle current vanishes. The Landau 
or energy frame, instead,  is the frame of observers with four-velocity 
$u^a_\mathrm{(L)} \neq u^a$ moving with the energy flow. In this 
frame, Landau observers see a diffusive particle flow described by a 
current $q^a_\mathrm{(p)}$ but not an energy 
flow, since the heat current density $q^a_\mathrm{(L)}$ vanishes 
identically. For a perfect fluid, the Eckart (comoving) and the Landau 
(energy) frames coincide and both the heat and the particle diffusion 
currents are zero. 

We have shown that $ {\cal T} = 0 $ but 
$\mu=\sqrt{2X}\neq 0$ in the comoving frame of the effective fluid 
associated with a minimally coupled scalar field. Here we check 
explicitly that 
this fact does not contradict the vanishing of the diffusion current 
because the two terms in the right hand side of 
Eq.~(\ref{generalizedFick}) 
cancel each other out. We have
\be
\nabla_a\mu = \nabla_a \left( \sqrt{2X} \right) = \frac{\nabla_a 
X}{\sqrt{2X}}
\ee
and the spatial gradient of $\mu$ is
\begin{eqnarray}
h_{ab}\nabla^b \mu =
\left( g_{ab} + \frac{\nabla _a \phi \nabla _b  \phi}{2 X} \right) \, 
\frac{\nabla^b X}{\sqrt{2X}}
 =
\frac{\nabla_a X}{\sqrt{2X}} + 
\frac{\nabla _b \phi \nabla^b X}{(2 X)^{3/2}} \, \nabla _a \phi
 \,.
\end{eqnarray}
Adding to this quantity the acceleration term $\mu \, \dot{u}_a$ and 
using the expression~(\ref{acceleration}) of the acceleration yields
\begin{eqnarray}
h_{ac} \nabla^c \mu + \mu \, \dot{u}_a = 0 \,,
\end{eqnarray}
which ensures that there is no diffusive ``particle'' current in this 
frame in spite of the non-uniform chemical potential.

\subsection{Phantom fields}
\label{subsec:4}

A phantom scalar field is obtained from a canonical one by changing the 
sign with which $X$ appears in the Lagrangian,
\be
{\cal L}=-X-V(\phi) \,,
\ee
which changes the usual stress-energy tensor to
\be 
T_{ab} = -\nabla_a \phi \nabla_b \phi + \frac{1}{2} \, g_{ab} \nabla^c\phi 
\nabla_c \phi -  g_{ab} V 
\ee 
and the equation of motion to
\be
\Box \phi +V_\phi=0 \,.
\ee
For example, in FLRW cosmology (which is no doubt the main area of theoretical 
physics contemplating phantom fields)  $\phi=\phi(t)$ and $X>0$, together with 
$V(\phi) \geq 0$. The phantom equation of state parameter is then
\be
w \equiv \frac{P}{\rho} = \frac{ {\cal L}}{ -2X-{\cal L}} =  
\frac{-X-V}{-X+V} = -1 -\frac{2X}{V-X} 
\ee
and the requirement $\rho >0 $, equivalent to $  0< X< V$, then 
yields $w<-1$. 

According to the previous section, for a phantom scalar field the particle 
number density, temperature, and  chemical potential are
\begin{eqnarray}
n &=&- \sqrt{2X} <0 \,,\\
&&\nonumber\\
{\cal T} &=& 0 \,,\\
&&\nonumber\\
\mu &=& \sqrt{2X} >0 \,.
\end{eqnarray}
While ${\cal T}$ and $\mu$ remain the same as for non-phantom fields, the 
number density $n$ becomes negative for phantoms. 

The thermodynamics of phantom fields and, more in general, phantom fluids, 
has been discussed in many works, usually beginning with  assumptions 
different from ours and often assuming negative 
temperature 
(or entropy) and positive chemical potential (or entropy), or {\em 
vice-versa}, from the outset and often considering 
quantum fields \cite{Nojiri:2004pf, Gonzalez-Diaz:2004clm, Gonzalez-Diaz:2004wun, Lima:2004wf, Babichev:2005nc,Izquierdo:2005ku,Izquierdo:2006mt, Setare:2006rf, Buchmuller:2006em, Gong:2006sn, Gong:2006ma, deFreitasPacheco:2007ysm, Pereira:2008af}. 
Usually, these discussions are limited to FLRW cosmology, where 
 negative temperatures were speculated independently  
\cite{Klemm:2004mb}. It is difficult to compare all these different (and 
sometimes contradictory) scenarios and assumptions, 
and to make 
sense of their conclusions caused by such an exotic, and most likely  
unphysical, field as the phantom. Moreover, our analogy is restricted to 
classical scalar fields. However, the picture that we offer for 
nonminimally coupled scalars in GR seems more  
grounded in fluid physics than many scenarios in the literature, in the 
sense that 
temperature and chemical potential are well-defined, with ${\cal T}=0$ and 
$\mu>0$, but $n <0$. 
This feature is definitely unphysical, as are many of  the consequences of 
a phantom field permeating the universe, and we will not consider phantom 
fields further.

\section{Nonminimal coupling to matter and Einstein frame formulation of 
scalar-tensor gravity}
\label{sec:3}
\setcounter{equation}{0}

Let us consider now the situation in which the scalar field couples nonminimally to 
other forms of matter, which are described by the Lagrangian density ${\cal 
L}^\mathrm{(m)}$ through a coupling function $f(\phi)$ (this coupling is non-trivial 
if $f(\phi)\neq$~const.).  For simplicity, we restrict to the scalar field Lagrangian 
${\cal L}=X-V(\phi)$. The total Lagrangian density is then
\be
{\cal L}= X- V(\phi) +f(\phi) {\cal L}^\mathrm{(m)} \,.
\ee
The equation of motion of $\phi$ becomes
\be
\Box \phi =V_\phi - f_\phi {\cal L}^\mathrm{(m)} \,. \label{vaimo}
\ee
The extra term acts as a source of $\phi$, hence as a source of ``particles'' in the 
effective $\phi$-fluid. As a consequence, the 
stress-energy tensors of $\phi$ and of the  other  matter are not 
conserved ($\nabla^b T_{ab}^{(\phi)} \neq 0$, $\nabla^b T_{ab}^\mathrm{(m)} \neq 0$) but their 
sum is, $\nabla^b \left( T_{ab}^{(\phi)} + T_{ab}^\mathrm{(m)}\right)= 0$. The 
coupling term on the right-hand side of Eq.~(\ref{vaimo})  
acts in the same way as the potential $V(\phi)$, preventing the conservation of the 
``particle'' current density  $N^a = nu^a=\nabla^a\phi$ according to 
Eq.~(\ref{ncons}). Indeed, this extra term  breaks the shift 
invariance $\phi \to \phi +C$ of the scalar field Lagrangian ${\cal 
L}={\cal L}(X)$ in the absence of a potential $V(\phi)$ and prevents $N^a$ 
from being a conserved Noether current even when  $V(\phi) \equiv 0$.\\\\

Since in the Einstein frame the scalar couples minimally to gravity but 
nonminimally to matter, these considerations open up the possibility of 
discussing the Einstein frame formulation of the thermodynamics of 
scalar-tensor gravity, which has so far been developed in the Jordan frame 
\cite{Faraoni:2021lfc, Faraoni:2021jri,Giardino:2022sdv,Faraoni:2022jyd}.

First-order thermodynamics deals with theories that have (Jordan 
frame) action
\be
S_{\rm ST}=\frac{1}{16\pi} \int d^4x \sqrt{-g} \left[ \phi R 
-\frac{\omega(\phi )}{\phi} 
\, \nabla^c\phi \nabla_c\phi -V(\phi) \right]+S^{\rm (m)},
\ee
where the Brans-Dicke scalar $\phi>0$ is approximately the inverse of the 
effective gravitational coupling $G_\mathrm{eff}$ and $\omega(\phi)$ is 
the ``Brans-Dicke coupling''. The scalar contribution to the 
energy-momentum tensor can be cast in the form of an effective imperfect 
fluid \eqref{12} \cite{Faraoni:2018qdr}.
Applying Eckart’s first-order non-equilibrium thermodynamics to this fluid 
allows one to recover an effective temperature ${\cal T}$ (and thermal 
conductivity ${\cal K}$)
\be
\label{kt}
{\cal K} {\cal T} = \frac{ \sqrt{-\nabla^c \phi \nabla_c \phi}}{ 8 \pi 
\phi} \,.
\ee
This is nothing but a temperature relative to GR, which represents the 
${\cal KT}=0$ equilibrium state. In this way, one can depict a landscape 
of gravity theories, where different theories (or classes thereof) are 
identified by their temperature relative to equilibrium and obtain an 
understanding of how this equilibrium might be approached through a 
dissipation process. However, the Einstein frame could not be handled in 
this picture based on 
the notion of temperature. The alternative and complementary picture 
relying on the chemical potential that was developed in the previous 
sections, on the other hand, can fill the gap.
We switch from the Jordan to the Einstein frame by performing the 
well-known conformal transformation of the metric \cite{Bekenstein:1974sf}
\be
\label{conftrasf}
g_{ab} \to \tilde{g}_{ab} \equiv \phi \, g_{ab} 
\ee
and the scalar field redefinition $\phi \to \tilde{\phi}$ with
\be
d\tilde{\phi} = \sqrt{ \frac{| 2\omega+3|}{16\pi}} \, \frac{d\phi}{\phi} \,.
\ee
The action then reads
\be
\label{BDactionEF}
S_\text{EF} =  \int 
d^4x \sqrt{-g} \left[ \frac{ \tilde{R}}{16\pi} -\frac{1}{2} \, 
\tilde{g}^{ab} \, \nabla_a\tilde{\phi} \nabla_b\tilde{\phi} 
-U(\tilde{\phi})  
+\frac{{\cal L}^\text{(m)}}{\phi^2(\tilde{\phi})}\right] \,,
\ee
with
\be
U (\tilde{\phi}) = \frac{V(\phi)}{16\pi \phi^2}\left|_{\phi=\phi( 
\tilde{\phi}) }  \right. \,.
\ee
All Einstein frame variables $\left( \tilde{g}_{ab}, \tilde{\phi} 
\right)$ are denoted 
by a tilde. The Einstein frame field equations read
\begin{eqnarray}
\tilde{R}_{ab}-\frac{1}{2} \, \tilde{g}_{ab} \tilde{R}  &=& 
8\pi \left(  \mbox{e}^{- \sqrt{\frac{64\pi}{|2\omega+3|}} 
\, \tilde{\phi} } \, T_{ab}^{\rm (m)} + 
\nabla_a \tilde{\phi} \nabla_b
\tilde{\phi} \right.\nonumber\\
&&\nonumber\\
&\, & \left.
-\frac{1}{2} \, \tilde{g}_{ab}\, 
\tilde{g}^{cd} \nabla_c \tilde{\phi} \nabla_d \tilde{\phi} 
 - U(\tilde{\phi})  \, \tilde{g}_{ab} \right)   \,,\nonumber\\ 
&& \label{Eframefe}
\end{eqnarray}
\be
 \tilde{g}^{ab} \nabla_a  \nabla_b
\tilde{\phi} -\frac{dU}{d\tilde{\phi}} 
+8\, \sqrt{ \frac{\pi}{|2\omega+3|}} \,  \mbox{e}^{ 
-\sqrt{\frac{64\pi}{|2\omega+3|}}\, \tilde{\phi} } \, {\cal 
L}^{\rm (m)} = 0 \,.
\label{EframeKG}
\ee
The scalar contribution to the stress-energy tensor arising from this 
action is of course that of a perfect fluid \eqref{11}. However, this 
presents a puzzle for the first-order thermodynamics of scalar-tensor 
theories. The thermodynamical formalism based on the temperature 
description is not 
suitable for a perfect fluid, since all imperfect fluid quantities vanish 
and the theory becomes trivial. This means that the approach to 
equilibrium cannot be studied. The formalism based on temperature only 
works for gravitational theories in representations where an effective 
imperfect fluid description can be found, which is possible only if the 
scalar is directly coupled to $R$ in the action. These considerations 
relate to the well-known but hard-to-tackle problem of the ambiguity that 
arises in distinguishing between ``gravitational'' and ``matter'' degrees 
of freedom whenever we switch representation through a conformal 
transformation \cite{Sotiriou:2007zu}.

However, the notion of chemical potential comes to the rescue. Although 
the temperature ${\cal T}$ of the Einstein frame scalar field effective 
fluid is zero, according to the previous sections, the chemical potential 
$ \tilde{\mu} = \sqrt{ 2\tilde{X}}$ is not. Now the scalar field 
$\tilde{\phi}$ has gravitational nature and is always present in 
spacetime, that is, one cannot decide to set it to zero or replace it with 
other forms of matter. The state of diffusive equilibrium corresponds to 
$\tilde{\mu}=0$ and $\tilde{\phi}=$~const., but this condition 
automatically recovers GR (possibly, with a cosmological constant if 
a potential for the scalar field is present),  as a limiting case given 
that a timelike gradient for the scalar field 
represents our starting assumption for this analogy.  This result goes 
hand-in-hand with that of 
first-order thermodynamics of scalar-tensor gravity formulated in the 
Jordan frame, where GR is the zero-temperature state of equilibrium 
\cite{Faraoni:2021lfc, Faraoni:2021jri, Giusti:2021sku,Giardino:2022sdv}. 
In the Einstein frame, instead, $\tilde{{\cal K}}\tilde{{\cal T}}$ is identically zero but 
GR is the state of equilibrium of scalar-tensor gravity corresponding to 
vanishing chemical potential $\tilde{\mu}=\sqrt{2\tilde{X}}$.  Therefore, the 
thermodynamical picture based on the chemical potential solves the issue 
and an understanding of the approach to equilibrium even for theories 
described by perfect fluids is possible. We would argue that the 
simplicity of the argument adds to the first-order thermodynamics of 
scalar-tensor gravity instead of detracting from it.

\section{Conclusions}
\label{sec:4}
\setcounter{equation}{0}

We have begun our discussion with minimally coupled scalar fields in GR 
and have corrected the current view of the analogy between these fields 
and effective perfect fluids with regard to temperature and chemical 
potential. This new view has allowed us to reformulate the first-order 
thermodynamics of scalar-tensor gravity as seen from the Einstein 
conformal frame, which was not possible earlier. Hence, we conclude this 
work from the broader view of the equivalent fluid of a scalar field 
nonminimally coupled to $R$ in scalar-tensor gravity. Two main conclusions 
emerge.

First, if the scalar field $\phi$ described by a Lagrangian 
density ${\cal L}\left( 
\phi, X \right)$ couples nonminimally with the Ricci 
scalar in the Jordan frame description of scalar-tensor gravity, the 
equivalent fluid has a nonvanishing 
temperature defined in \cite{Faraoni:2018qdr, Faraoni:2021lfc, 
Faraoni:2021jri, Giusti:2021sku} (exceptions are theories in which the 
scalar field is non-dynamical \cite{Faraoni:2022doe}).

Second, in GR a scalar field coupled minimally to $R$ (but 
possibly nonminimally to other forms of matter) has zero temperature 
${\cal T}$ but nonvanishing chemical potential $\mu=\sqrt{2X} $. However, 
in the comoving (or Eckart) frame, no diffusive flux of 
``$\phi$-particles'' is visible because this frame follows the total 
motion of this effective fluid. This situation includes the Einstein frame 
description of scalar-tensor gravity and allows one to establish that GR 
is the state of diffusive equilibrium ({\em i.e.}, $\mu \equiv 0$) of 
scalar-tensor gravity formulated in the Einstein frame. The previous 
approaches to first-order thermodynamics of scalar-tensor and Horndeski 
gravity \cite{Faraoni:2021lfc, Faraoni:2021jri, 
Giusti:2021sku,Giardino:2022sdv,Faraoni:2022jyd} were based on the Jordan 
frame description and on the {\em temperature} of the $\phi$-fluid and 
were thus unable to deal with the Einstein frame. Realizing that scalar 
fields minimally coupled to the curvature should have zero temperature but non-zero {\em chemical potential} is the key to resolve that conundrum.

This work is limited to situations in which there are no derivative 
couplings of the scalar and no second derivatives of $\phi$ in the 
Lagrangian, while theories with ${\cal L}={\cal L}\left( \phi, X, \Box 
\phi\right) $ are the subject of much attention in the literature.  Other 
questions arise naturally: what is the Landau frame for the fluid 
equivalent of a nonminimally coupled scalar field?  The discussion in the 
literature thus far has exclusively been based on the comoving (Eckart) 
frame, but the choice of the Landau frame is advantageous in the analysis 
of relativistic fluids in nuclear collisions 
\cite{BRAHMS:2004adc,PHOBOS:2004zne, STAR:2005gfr,PHENIX:2004vcz, 
Monnai:2019jkc} and may disclose unexpected view of scalar fields as well. 
Future work will focus on these questions.

\begin{acknowledgments} 

V.F. and A.G. are grateful to Alexander Vikman for pointing out 
Refs.~\cite{Deffayet:2010qz, LimSawickiVikman10, 
Pujolas:2011he,Mirzagholi:2014ifa} and for suggesting the chemical 
potential approach to scalar-tensor thermodynamics.  This work is 
supported, in part, by the Natural Sciences \& Engineering Research 
Council of Canada (Grant 2016-03803 to V.F.). S.G. thanks Jean-Luc Lehners 
at AEI Potsdam for hospitality. A.G.~is supported by the European Union's 
Horizon 2020 research and innovation programme under the Marie 
Sk\l{}odowska-Curie Actions (grant agreement No.~895648--CosmoDEC). The 
work of A.G~has also been carried out in the framework of the activities 
of the Italian National Group of Mathematical Physics [Gruppo Nazionale 
per la Fisica Matematica (GNFM), Istituto Nazionale di Alta Matematica 
(INdAM)].

\end{acknowledgments}

\bigskip
\appendix
\section{$s/n=$~const. along perfect fluid lines}
\label{AppendixA}
\renewcommand{\theequation}{A.\arabic{equation}}

Here we reproduce a standard result of perfect fluids stating that the 
entropy per particle $s/n$ is constant along the flow lines of a perfect 
fluid in which entropy and particle number are conserved. As a 
consequence, neighbouring fluid lines do not exchange entropy per particle 
and different values of $s/n$ distinguish different flow lines.

Consider a flow line with four-tangent $u^a$ parametrized by the proper 
time $\tau$; we have 
\be
\frac{d}{d\tau} \left( \frac{s}{n} \right) = \frac{1}{n} \, 
\frac{ds}{d\tau} - 
\frac{s}{n^2}\, \frac{dn}{d\tau}  \,.
\ee
Conservation of particle number gives
\be
0=\nabla_c N^c = \nabla_c \left( nu^c \right) =n\nabla_c u^c + u^c 
\nabla_c \, n 
\ee
and
\be
\frac{dn}{d\tau} \equiv u^c\nabla_c n =- n \nabla_c u^c \,.
\ee
Likewise, conservation of entropy for a perfect fluid without 
dissipation yields
\be
0=\nabla_c s^c = \nabla_c \left( su^c \right) =s\nabla_c u^c + u^c 
\nabla_c \, s 
\ee
and
\be
\frac{dn}{d\tau} =- s \nabla_c u^c \,.
\ee
Then we have 
\be
\frac{d}{d\tau} \left( \frac{s}{n} \right) =\frac{1}{n} \left( 
\frac{ds}{d\tau} -\frac{s}{n} \, \frac{dn}{d\tau} \right) =\frac{1}{n} 
\left( -s\nabla_c u^c + \frac{s}{n} \, n \nabla_a u^a \right) =0 \,.
\ee

For the effective $\phi$-fluid, $s/n$ cannot be identified with the scalar 
field $\phi$, as done in previous literature. Indeed, as shown above, 
$s/n$ is constant 
along  fluid lines (this is certainly the case also for the effective 
$\phi$-fluid if ${\cal L}_{\phi}=0$), while $\phi $ necessarily changes 
along the flow lines. In fact, in general, $\phi$ depends on the proper 
time $\tau$ along the flow lines, as well as on the spatial coordinates of 
the 3-space orthogonal to these flow lines. Indeed, in applications to 
FLRW cosmology (the main purpose of 
Refs.~\cite{Deffayet:2010qz,LimSawickiVikman10,Pujolas:2011he,Mirzagholi:2014ifa}), 
$\phi$ depends {\em only} on $\tau$, which coincides with the FLRW 
comoving time. The situation in which $\phi$ depends only on $s/n$ and, in 
particular, the identification $\phi=s/n$ would mean that there is a 
coordinate system in which the gradient $\nabla_a \phi $ is purely 
spatial, then 
the latter is spacelike and cannot be timelike, which is instead 
essential for identifying~(\ref{4-velocity}) with the effective fluid 
four-velocity and constructing the effective fluid description of the 
scalar field.

\end{document}